\documentclass{pasj00}
\draft

\begin{document}
\SetRunningHead{S. Arimura \etal}
{Wide-Area Mapping of 155 Micron Continuum Emission from 
    the Orion Molecular Cloud Complex}

\Received{2003/07/06}
\Accepted{2003/12/04}

\title{Wide-Area Mapping of 155 Micron Continuum Emission from 
       the Orion Molecular Cloud Complex}

\author{
   Seikoh              \textsc{Arimura,}\altaffilmark{1,2}
   Hiroshi             \textsc{Shibai,}\altaffilmark{1}
   Takafumi            \textsc{Teshima,}\altaffilmark{1}
   Takao               \textsc{Nakagawa,}\altaffilmark{3}\\
   Masanao             \textsc{Narita,}\altaffilmark{3}
   Shin'itirou         \textsc{Makiuti,}\altaffilmark{3}
   Yasuo               \textsc{Doi,}\altaffilmark{4}
   Ram Prakash  	   \textsc{Verma,}\altaffilmark{5}\\
   Swarna Kanti        \textsc{Ghosh,}\altaffilmark{5} 
   Thinnian Naganathan \textsc{Rengarajan,}\altaffilmark{5,6}
   Makoto              \textsc{Tanaka,}\altaffilmark{7}
   \and
   Haruyuki            \textsc{Okuda}\altaffilmark{8}
}   

\altaffiltext{1}{
Department of Astrophysics, School of Science, Nagoya University, \\
Furo-cho, Chikusa-ku, Nagoya 464-8602
}  
\email{arimura@u.phys.nagoya-u.ac.jp}
\email{shibai@u.phys.nagoya-u.ac.jp}
\altaffiltext{2}{
Institute of Space and Astronautical Science, \\
Japan Aerospace Exploration Agency (JAXA), \\
2-1-1 Sengen, Tsukuba, Ibaraki 305-8505}  
\email{arimura.seikoh@jaxa.jp}
\altaffiltext{3}{
Institute of Space and Astronautical Science, \\
Japan Aerospace Exploration Agency (JAXA), \\
3-1-1 Yoshinodai, Sagamihara, Kanagawa 229-8510}  
\altaffiltext{4}{
College of Arts and Sciences, The University of Tokyo, \\
Komada 3-8-1, Meguro, Tokyo 153-8902}  
\altaffiltext{5}{
Tata Institute of Fundamental Research, 
Bombay 400 005, India} 
\altaffiltext{6}{
Instituto Nacional de Astrofisica Optica y Electronica, \\
Puebla 72840, Mexico} 
\altaffiltext{7}{
NEC Corp., Shimonumabu, Nakahara, Kawasaki 211-8666}  
\altaffiltext{8}{Gunma Astronomical Observatory, \\
Nakayama 6860-86, Takayama, Agatsuma, Gunma 377-0702}



\maketitle

\begin{abstract}
We present the results of a wide-area mapping of the far-infrared continuum emission toward the Orion complex by using a Japanese balloon-borne telescope. The 155-$\mu$m continuum emission was detected over a region of 1.5 deg$^2$ around the KL nebula with $\timeform{3'}$ resolution similar to that of the IRAS 100-$\mu$m map.
Assuming a single-temperature model of the thermal equilibrium dust, maps of the temperature and the optical depth were derived from the 155 $\mu$m intensity and the IRAS 100 $\mu$m intensity. The derived dust temperature is 5 -- 15 K lower and the derived dust optical thickness were derived from the 155-$\mu$m intensity and the IRAS 100-$\mu$m intensity.
The derived dust temperature is 5 -- 15 K lower and the derived dust optical depth is 5 -- 300 times larger than those derived from the IRAS 60 and 100-$\mu$m intensities due to the significant contribution of the statistically heated very small grains to the IRAS 60-$\mu$m intensity.
The optical-thickness distribution shows a filamentary dust ridge that has a $\timeform{1D.5}$ extent in the north -- south direction and well resembles the Integral-Shaped Filament (ISF) molecular gas distribution.
The gas-to-dust ratio derived from the CO molecular gas distribution along the ISF is in the range 30 -- 200, which may be interpreted as being an effect of CO depletion due to the photodissociation and/or the freezing on dust grains.
\end{abstract}

\section{Introduction}
The Orion region at a distance of 470 pc \citep{Brown1994}, is the nearest active star-forming region containing massive stars. It contains a massive molecular cloud complex, bright H\emissiontype{II} regions, and many young stellar objects.
Many studies of this region as a representative case of a massive-star-forming region have been made and have provided insight into the massive-star-formation process (see reviews by \cite{Goudis1982, Genzel1989, O'Dell2001}). Even now, it is the most important region for investigating the massive-star-formation phenomena.
M 42 is the brightest H\emissiontype{II} region in this region, and is excited by the Trapezium star cluster. 
A dense molecular cloud complex lies behind the H\emissiontype{II} region. In the densest part of this molecular could (OMC-1) are the Orion KL and BN objects \citep{KL, BN}. They are embedded deeply in OMC-1, and energize the surrounding molecular gas of OMC-1. Two other dense molecular clouds, OMC-2 and -3, lie along the dense molecular ridge north of Orion KL.
This dense ridge has a filamentary structure, $\timeform{1.5D}$ long. From the apparent shape seen in the ${}^{13}$CO high-resolution map by \citet{Bally1987}, it was named the Integral-Shaped Filament (hereafter ISF). The cloud cores of OMC-1, -2, and -3 lie on the dense filamentary feature.
At the northernmost position of ISF, there is a H\emissiontype{II} region, NGC 1977, and the south of it, there are small H\emissiontype{II} regions, L 1641-N and NGC 1999.
A H\emissiontype{II} region, M42 containing the Orion KL, lies on the middle of the ISF.
The ISF has so far been observed and identified in various molecular lines of CO (J = 2--1) by \citet{Sakamoto1994}, ${}^{13}$CO (J = 1--0) by \citet{Nagahama1998}, C${}^{18}$O (J = 1--0) by \citet{Dutrey1991}, NH${}_3$ by \citet{Cesanori1994}, and CS by \citet{Tatematsu1993}, and in the neutral atomic carbon [C\emissiontype{I}] line by \citet{Ikeda2002}.

The continuum emission in the far-infrared and sub-millimeter wavelength regions is radiated by interstellar dust grains that are mixed with the molecular gas.
Thanks to recent progress in instrumental techniques in this wavelength region, relatively wide areas of the sky have been mapped with ground-based telescopes \citep{Chini1997, Lis1998, Johnstone1999} and with balloon-borne telescopes \citep{Ristorcelli1998, Mookerjea2000-a, Dupac2001}.
Previous investigations successfully revealed the dust distribution of this region, but covered only a part of the ISF-OMC-1 and its vicinity, OMC-2, and OMC-3.
\citet{Mookerjea2000-a} mapped this region at 138 and 205 $\mu$m with a spatial resolution of $\timeform{1.5'}$ using a one-meter balloon-borne telescope, and showed that a dust ridge extends from Orion KL to the north, tracing the molecular ridge.
\citet{Ristorcelli1998, Dupac2001} obtained wider-area maps at 200, 260, 360, and 580 $\mu$m with the large balloon-borne telescope, PRONAOUS. The observed area was from Orion KL to the north. 
Their observations revealed several cold condensations to the west of OMC-1, which may be pre-collapse phase cloud cores.
\citet{Johnstone1999} mapped the southern ridge with SCUBA of JCMT, and showed that the dust ridge really extends along the ISF at least up to $\timeform{30'}$ from Orion KL.
However, the observed area ($\timeform{50'} \ \times \ \timeform{10'}$) did not cover the whole of the ISF. 
It is of great interest to map the interstellar dust distribution over the whole region of the ISF. 
As is well known, IRAS completely mapped this region at four infrared bands of 12, 25, 60, and 100 $\mu$mm.
By using the IRAS 60 and 100 $\mu$m maps, \citet{Bally1991} derived the maps of dust temperature and the dust mass.
However, they could not identify the ISF-like structure in the derived maps. 
The main reason for this is the unsuitability of the IRAS 60 $\mu$m band for the study of a cold dust region like the ISF, because in this band, the emission from stochastically-heated, very small grains significantly contaminates the thermal emission from the large grains at a steady state temperature (e.g. \cite{Nagata2002}).
Moreover, as is mentioned later, the IRAS 60 and 100 $\mu$m band signals are saturated at and near the Orion KL position and have a spurious pattern due to the insufficient baffle design.
COBE/DIRBE had achieved highly reliable, well-calibrated intensity maps at various infrared bands for the whole sky. The 100, 140, and 240 $\mu$m intensity maps of DIRBE can be used for deriving the dust temperature and mass distribution. However, the spatial resolution of $\timeform{0.7D}$ of the DIRBE maps is insufficient for resolving the dust distribution in the Orion region \citep{Wall1996}.
The aim of the present work is to observe a very large area covering the whole ISF at wavelengths between 100 and 200 $\mu$m, where the interstellar dust (large grains) has its peak emission, with a spatial resolution similar to that of IRAS 100 $\mu$m band in order to reveal the dust temperature and mass distribution and, consequently, the energy source of individual parts of molecular clouds in this region.

The instrumentation and observation are described in section 2; the data reduction, analysis, and calibration are explained in section 3. 
The intensity map and the temperature distribution are presented in section 4.
In section 5, the column-density distribution of the dust grain is derived, and the heating sources for this region are discussed.

\section{Instrumentation and Observation}
\subsection{Instrumentation}
A balloon-borne telescope, \textit{FIRBE (\underline{F}ar-\underline{I}nfra\underline{r}ed \underline{B}alloon-Borne \underline{E}xperiment)}, was newly developed to observe the far-infrared continuum emission over a wide sky area \citep{Arimura2000, Doi2000-a, Shibai2002}.
The telescope was designed for suppressing the thermal emission from non-cooled devices, such as mirrors and structural components. The primary mirror is a F/2.1 off-axis paraboloid of 50 cm aperture with no other warm mirror or structural part. The converging beam after reflection by the primary mirror is introduced into a cryostat through a thin polypropylene vacuum window.
A two-dimensional stressed Ge:Ga array detector \citep{Doi2000-b} is installed in the cryostat with a cold optical system that converts the beam F ratio and blocks the light coming from the beam with a Lyot-stop. The detector array is cooled down to 1.8 K with super-fluid helium. 
The array consists of 4 $\times$ 8 pixels, each with a field of view of $\timeform{1'.5}$.
The total field of view is $\timeform{12'}$ in the elevation direction by $\timeform{6'}$ in the cross-elevation direction.
The spectral-response curve is shown in figure 1. This curve represents the system performance (mean of all 32 pixels), including the detector response, the filter response, and all other effects. The pixel-to-pixel variation is negligible. The spectral response curves of the IRAS 100 $\mu$m band and the COBE/DIRBE 140 $\mu$m band are presented in this figure.
The effective wavelength and effective bandwidth are 155 $\pm$ 1.5 $\mu$m and 33.8 $\pm$ 2 $\mu$m, respectively, for a modified Plank function spectrum with a $\lambda^{-2}$ emissivity law for dust temperatures from 15 to 30 K.

\subsection{Observation}
The balloon launch of FIRBE was undertaken as an international collaboration project between Nagoya University and Tata Institute of Fundamental Research (TIFR) of India at the National Balloon Facility of India. This facility is located at $\timeform{17D.47}$N, $\timeform{78D.57}$E in Hyderabad, India.
The launch date and time was 16:30 (UT) on 2000 December 16. The flight number was 448.
The data presented in this paper were obtained during a period of two hours from 21:30 to 23:30 (UT). 
The detector noise was nearly equal to the background photon noise, as expected from pre-flight measurements.
The attitude control system is an alt-azimuth mounting; the gondola payload is oriented in the azimuthal angle by a reaction wheel mechanism, and the telescope is driven around the elevation axis of the gondola.
A wide-area survey was executed by repeating back-and-forth scans in azimuth with periodic stepwise control in elevation.
The scanning span was $\timeform{8D}$ in azimuth.
The signal from each detector pixel was sampled every 1/32 seconds, which corresponds to 12" in the scanning direction.
The surveyed sky area was over 50 deg$^2$. 
The large-area mapping could detect significant far-infrared emissions around Orion KL with an extent of over 1.5 deg$^2$ and the neighborhood around NGC 2023 and 2024 in Orion B (south).

\section{Data Reduction}
A post-observational attitude reconstruction was made using the following procedure.
The primary attitude sensor was a 3-axis gyroscope. 
The resolution of this devise was better than $\timeform{3"}$, and the speed of the data sampling rate (32 Hz) was fast enough for attitude reconstruction.
However, the gyroscope has a relatively large, long-term drift on the order of $\timeform{1'}$ per 5 minutes, which must be corrected to achieve the required attitude accuracy over one-hour observation period. 
Therefore, auxiliary attitude sensors, a geomagnetic aspect sensor and a star-field camera, were employed mainly for absolute attitude determinations.
The final positional accuracy was better than $\timeform{3'}$ over the whole mapped area.
The instrumental beam size was $\timeform{2'.9}$ (FWHM) circular according to the laboratory measurement before flight, whereas the expected beam size was $\timeform{2'.0}$, considering the diffraction limit of $\timeform{1'.3}$ and the pixels size of $\timeform{1'.5}$. 
This degradation is possibly due to an insufficient optical adjustment.
A binning grid size of $\timeform{3'}$ was adopted for the observed intensity map. 
Since the primary purpose of the present work is to study the extended emission, this binning size is reasonable.

The raw detector signals were processed as follows. 
The raw detector output is dominated by the thermal emission from the instruments and the earth's atmosphere in the line of sight.
At first, in order to subtract the contribution by these non-astronomical signals, the sky area positions having the flux density less than 100 MJy sr$^{-1}$ in the IRAS 100 $\mu$m band were defined as the zero-intensity positions.
The signal above the zero-intensity level was extracted as astronomical emission.
Next, in order to correct for any differences among all detector pixels, the flat-fielding method was applied by using the output signal for the whole mapped area.
Thirdly, the COBE/DIRBE 140 $\mu$m intensity was used for determining the true zero-intensity level and the intensity scale.
For a precise comparison of both intensities, the beam sizes were convolved with a Gaussian beam profile of $\timeform{1D.4}$ (FWHM), which is twice as large as the COBE/DIRBE beam size.
The resultant intensity scale and the offset intensity of the zero-level from the true zero-flux level were $1085 \pm 10$ MJy sr$^{-1}$ mV$^{-1}$ and $\sim 80.8 \pm 2$ MJy sr$^{-1}$, respectively.
The reason why the offset intensity is a non-zero positive value is that a diffuse emission extends to the zero-intensity positions.
The absolute uncertainty of the FIRBE map is dominated by the COBE/DIRBE calibration uncertainty of 10\% (COBE/DIRBE Explanatory Supplement).

\section{Result}

\subsection{Intensity Map}
Figure 3a shows an intensity map of far-infrared continuum intensity at 155 $\mu$m obtained by the present work.
The lowest contour indicates 352 MJy sr$^{-1}$, corresponding to the 2.6 $\sigma$ noise level.
The intensity peak coincides with the Orion KL position, where the intensity is 71080 $\pm$ 8000 MJy sr$^{-1}$, corresponding to a flux value of 42514 $\pm$ 4785 Jy in the $\phi \ \timeform{3'}$ beam.
No intense and extended emission around the KL nebula is seen, except in the eastern region, where smooth extended emission exists.
The spatial distributions around the Orion KL and the northern filamentary structure are consistent with 138 and 205 $\mu$m maps with higher $\timeform{1'.5}$ spatial resolution presented by \citet{Mookerjea2000-a}.
However, the extended diffuse emission detected by the present work is not seen in their maps.
The reasons for this are possibly that their sensitivity is 2.5 times lower than that of the present work (136 MJy sr$^{-1}$ 1$\sigma$) and the chopped observation with AC-coupled bolometer used by them is not sensitive to smooth extended emission.
By using the PRONAOS telescope, \citet{Dupac2001} obtained intensity maps in four photometric bands (200, 260, 360, and 580 $\mu$m) with an angular resolution of $\timeform{2'}$ to $\timeform{3'.6}$ in the area of the northern extended region containing the Orion KL nebula.
The mapped region was with relative right ascension from $\timeform{-4m}$ to $\timeform{+2m}$ and relative declination from $\timeform{+30'}$ to $\timeform{-10'}$ from Orion KL position.
The Noise Equivalent Intensity (NEI) of the present work is 54 MJy sr$^{-1}$ Hz$^{-1/2}$ at 16 Hz, and is 2.5 times larger than that of the PRONAOUS 200 $\mu$m band (22 MJy sr$^{-1}$ Hz$^{-1/2}$).
In regions that have intensity higher than 1 GJy sr$^{-1}$, the FIRBE 155 $\mu$m map is consistent with the PRONAOUS maps.
However, in regions of intensity of less than 1 GJy sr$^{-1}$, the PRONAOUS values are significantly lower as compared to the present 155 $\mu$m map. 
This inconsistency can again be attributed to the fact that an AC-coupled bolometer sensor was used for PRONAOUS, whereas the present work used a DC-coupled Ge:Ga photoconductor.
An AC-coupled bolometer is generally insensitive for slowly changing, extended emission.
\citet{Dupac2001} have found four dust concentrations (Cloud 1, 2, 3, and 4) to the west of the Orion KL nebula. In the FIRBE 155 $\mu$m map, only two sources, Cloud 1 and Cloud 2 can be seen as sources above the lowest contour level; the expected intensities of Cloud 3 and Cloud 4 are below the lowest contour level plotted.

The IRAS 100 $\mu$m intensity map of the same region is shown in figure 3b for a comparison.
The intensity scale of the IRAS 100 $\mu$m was corrected for the IRAS -- COBE transformation gain of 0.72 so as to match the COBE/DIRBE calibration (COBE/DIRBE Explanatory Supplement).
\citet{Bally1991} indicated that the IRAS 60 and 100 $\mu$m detectors must be saturated near the Orion KL, because the maps are unrealistically flat there.
In addition to the saturation, a spurious emission pattern can be seen in the IRAS 100 $\mu$m map. 
This spurious pattern shows a ring-like structure having a radius of $\timeform{50'}$ -- $\timeform{75'}$ centered on Orion KL.
Except for the saturation region and the spurious-patterned region, the FIRBE 155 $\mu$m map generally shows the same characteristics as the IRAS 100 $\mu$m map regarding the following points:
(1) the northern second peak associated with IRAS 05327$-$0457,
(2) an arc-shaped structure surrounding the northern H\emissiontype{II} region, NGC 1977,
(3) a filamentary ridge from Orion KL to IRAS 05327$-$0457,
(4)a filamentary ridge from Orion KL to the southwestern source, IRAS 05333$-$0543,
(5) a wide plateau region on the west of Orion KL, and
(6) southwest wide-diffuse region at an angular distance around $\timeform{30'}$ form Orion KL.
However, one can also see several differences between the FIRBE 155 $\mu$m map and the IRAS 100 $\mu$m map.
The most remarkable structure in the present map is a filamentary ridge extending along the ISF to the south.
This extended ridge is connected to the south-southeastern diffuse emission region surrounding the small H\emissiontype{II} regions, L 1641-N and NGC 1999.
Additionally, a weak emission component extends to the northwest direction from Orion KL to the position of R.A. = $\timeform{5h33.4m}$ -- $\timeform{5h34.8m}$ , DEC. = $\timeform{-4D45'}$ -- $\timeform{-5D10'}$.

\subsection{Dust Temperature and Optical Depth}
By assuming a single-temperature dust component model with the $\lambda^{-2}$ emissivity law, the dust temperature and optical depth can be derived from the 155 $\mu$m intensity of the present work ($I_{155}$) and the IRAS 100 $\mu$m intensity  ($I_{100}$). 
The dust grains responsible for the far-infrared intensity at wavelengths equal to or longer than 100 $\mu$m are large grains at a steady-state temperature.
Hereafter, this temperature is denoted as $T_{\mathrm{LG}}$.
It may be noted that because both data were taken with wide photometric bands, color-correction based on the spectral response curve shown in figure 1 is needed. 
The color-correction factor, $K_{\mathrm{\lambda}}$, depends on the spectral response of photometric band, dust temperature, and an emissivity law.
Consequently, the dust temperature ($T_{\mathrm{LG}}$) and optical depth at 100 $\mu$m ($\tau_{100}$) are estimated from the following equations:
\begin{eqnarray}
I_{100}/K_{100}(T_{\mathrm{LG}}) &=& \Bigl[ 1- \exp(-\tau_{100}) \Bigr] B_{100}(T_{\mathrm{LG}}),\\
I_{155}/K_{155}(T_{\mathrm{LG}}) &=& 
\Bigl\{ 1- \exp \bigl[ -\tau_{100} ({}^{\underline{100}}_{155})^2 \bigr] \Bigr\} B_{155}(T_{\mathrm{LG}})\\
\end{eqnarray}
where $B_{100}$ and $B_{155}$ are the Plank functions at 100 and 155 $\mu$mm.
The temperature and the optical depth of the dust grain are primarily important parameters for investigating the interstellar matter. From the two values, representative values of the interstellar radiation field (ISRF), the total luminosity of the dust emission, and the column density of the interstellar matter can be derived for each line of sight.
The dust temperature map and the gas column density map, thus obtained, are shown in figure 4a and in figure 6a, respectively. 
Figure 4b shows the dust temperature map derived from the IRAS 60 and 100 $\mu$m intensities in the same manner.
The total column density (gas + dust) shown in figure 6a was derived assuming a gas/dust ratio of 100.

A square field of $\timeform{15'} \times \timeform{15'}$ centering at Orion KL is excluded because the IRAS 100 $\mu$m intensity is saturated there.
From a comparison of the two temperature maps of figure 4, two differences are found:
(1) the dust temperature T(60/100) derived from the two IRAS bands is, on an average, 10 Kelvin higher than $T_{\mathrm{LG}}$ and 
(2) the T(60/100) map (figure 4b) shows the same spurious-pattern as seen in the IRAS 100 $\mu$m map.
\citet{Mookerjea2000-a} presented a dust temperature map derived from their 138 and 205 $\mu$m intensities around Orion KL neighborhood and northern ridge.
However, because their temperature map covers too small an area, a comparison with the present work is not meaningful.

\section{Discussion}

\subsection{Warm Dust Region Surrounding the H\emissiontype{II} Region}
In this section, the warm dust region surrounding the H\emissiontype{II} region, M 42, is discussed.
As described in section 4, two temperature maps are shown in figure 4a and figure 4b. They were derived from the intensity ratio of the IRAS 100 $\mu$m to the FIRBE 155 $\mu$m and that of the IRAS 60 $\mu$m to the IRAS 100 $\mu$mm, respectively, using equations (1) and (2).
It can be seen in both figures that warm dust is associated with the two H\emissiontype{II} regions, M 42, and NGC 1977. 
Apparently, the warm dust distribution is different from that of the CO molecular gas region.

First, we discuss the warm dust region associated with M 42.
Figure 5 shows a schematic diagram for the definition of the three regions around M 42 superposed on the red band image.
From figure 4a, it can be seen that an elliptical-shaped region where the dust is warmer than 20 K, defined as Warm Dust Region having a size of $\timeform{50'}$ in the E -- W direction. 
In addition, a narrower Cold Dust Region of less than 20 K exists from the KL nebula to the south.
This Cold Dust Region is discussed later.
Except for this region, Warm Dust Region has an extent similar to that of the region hotter than 28 K in figure 4b [the T(60/100) dust temperature map] and the optical extent in figure 4c (the red band image of Digital Sky Survey).
This fact suggests that the far-infrared emission in this Warm Dust Region is possibly from the dust grains coexisting with the neutral hydrogen gas just outside of the ionization front, not from the dust grains in the dense molecular gas. 
The neutral gas there is considered to be heated by the Trapezium star clusters lying in front of the dense cloud, and the dense molecular gas to be heated by deeply embedded sources, such as Orion KL. 
Therefore, except for the brightest region around Orion KL, the energy source of the far-infrared emission in this warm dust region is mainly provided by the Trapezium cluster, not by the Orion KL.
\citet{Bally1991} arrived at the same conclusion from the similarity between the dust temperature distribution derived from the IRAS 60 and 100 $\mu$m and the optical image of the Red band. 
The IRAS 60 $\mu$m band cannot be attributed wholly to the thermal emission from the large grain. 
The present work has confirmed the same conclusion from more reliable observational data.
In contrast to the Warm Dust Region, the dust temperature is low in the Cold Dust Region, where the surface brightness is high in the Red band image.
This implies that the column density of the cold dust is very high in this line of sight. Actually, the integrated line intensity of ${}^{13}$CO (figure 6b) shows a similar distribution with the narrow Cold Dust Region in figure 4a around this area.
\citet{Dutrey1991} found that the dense molecular gas exists in the Cold Dust Region by observing the C${}^{18}$O molecular line, which is sensitive to a gas density of about $10^4$ cc$^{-1}$.

Secondly, we discuss the other small region near Orion KL. 
In the temperature map derived from the IRAS 60 and 100 $\mu$m (figure 4b), a region with a dust temperature of more than 35 K, described as `High' Temperature Region in figure 5, can be seen in the western neighborhood from Orion KL. 
However, this `High' Temperature Region cannot be recognized in the temperature map derived from the FIRBE 155 $\mu$m and the IRAS 100 $\mu$m (figure 4a).
In this region, the red band brightness is fully saturated (figure 4c).
This fact means that the H${\alpha}$ line is predominantly strong in the red band.
The extent of this saturated region in the red band image is similar to that of the ionized gas traced by radio continuum emission observed by \citet{Subrahmanyan2001} (see figure 4c).
The extent of radio continuum emission is drawn in plane white color in figure 4c.
The `High' Temperature Region is considered to be the M 42 H\emissiontype{II} region, itself.
Since the temperature map in figure 4a does not show this `High' Temperature Region. It may be attributed to the `higher' intensity in the IRAS 60 $\mu$m band. 
Generally, the 60 $\mu$m intensity is higher than that extrapolated from the longer wavelength SED \citep{Boulanger1996, Shibai1999}. 
However, the difference between figure 4a and figure 4b suggests to us that the excess of the 60 $\mu$m intensity is much larger than that generally seen in the far-infrared SED in the galactic plane.
Since the optical depth is much less than unity, as described in the next subsection, optically thin case will be assumed throughout of this paper.

In order to investigate the difference between the `High' Temperature Region and the Warm Dust Region, the ratios of the mean intensities in these two regions are listed in table 1. 
The dust temperature from the FIRBE 155 $\mu$m band and the IRAS 100 $\mu$m band ranges from 21.8 to 30.8 K in `High' Temperature Region. 
Outside of this region, the reference region is defined to be the region where the dust temperature is in the same range. 
The intensity ratio is the ratio of the mean intensity in the `High' Temperature Region to the mean intensity in the reference region of Warm Dust Region.
The mean intensity ratio is almost constant for the IRAS 100 $\mu$m band and the FIRBE 155 $\mu$m band. 
This means that the dust temperatures of large grains are almost the same for the two regions, i.e. the Warmer Dust Region and the reference region in Warm Dust Region. Surprisingly, the IRAS 12 $\mu$m band has the same tendency. 
However, the ratios in the IRAS 25 $\mu$m and 60 $\mu$m bands are 40\% higher than others.
This tendency is consistent with the fact found by \citet{Okumura1999} for several H\emissiontype{II} regions in the galactic plane.
They examined the correlation between the ratios of $\nu I_\nu$ at 12, 25, and 60 $\mu$m bands of IRAS to the total far infrared intensity from the large grain 
$(I_{\mathrm{FIR}})$, and the strengh of the interstellar radiation field $(G_0)$, and found rather strong correlations over a wide range of $G_0$ from 1 to 100.
They showed that the $\nu I_\nu / I_{\mathrm{FIR}}$ ratio in the IRAS 12 $\mu$m intensity was almost constant.
This result indicated that the intensity of the IRAS 12 $\mu$m band emission was proportional to $I_{\mathrm{FIR}}$, which was also proportional to the product of the optical depth at 100 $\mu$m and $G_0$ $(I_{\mathrm{FIR}} \propto \tau_{100} \cdot G_0)$.
They suggested that the 12 $\mu$m emission was radiated mainly by the PAH-type large molecules through a fluorescence process and, thus, was proportional to $G_0$.
On the other hand, the $\nu I_\nu / I_{\mathrm{FIR}}$ ratios at the 25 and 60 $\mu$m bands were nearly proportional to $G_0$. This implies that the IRAS 25 and 60 $\mu$m intensities are proportional to the square of the strength of ISRF $(G_0^{\ \sim 2})$.
They attributed this tendency to the result of the transient heating of very small grains by double photon incidence by which the 25 and 60 $\mu$m intensities could be proportional to the square of $G_0$ under a possible condition. 
The $G_0$ value derived from the intensity ratio of the IRAS 100 $\mu$m to the FIRBE 155 $\mu$m in the `High' Temperature Region distributes in the range from $\sim$ 10 to 150 within their $\timeform{4'}$ beam size.
This result indicates that the strength of the UV radiation field of the `High' Temperature Region in the M 42 H\emissiontype{II} region has a similar strength to that of several H\emissiontype{II} regions in the galactic plane.
We find similar correlations that $\nu I_\nu /I_{\mathrm{FIR}}$ for the IRAS 12 $\mu$m band and those for the IRAS 25 and 60 $\mu$m bands are proportional to the power law of $G_0$.
Of course, the intensities at the IRAS 100 $\mu$m and the FIRBE 155 $\mu$m bands can be expected to be proportional to $I_{\mathrm{FIR}}$.
Therefore, by adopting the model proposed by \citet{Okumura1999}, it can be explained that the intensities at 25 and 60 $\mu$m are much more enhanced than those at 12, 100, and 155 $\mu$m in the M 42 H\emissiontype{II} region, where the ISRF must be much stronger than the general interstellar radiation field, and thus the significant higher temperature in `High' Temperature Region can not be recognized in the T(100/155) map shown in figure 4a.

\subsection{Integral-Shaped Filament and South-Southeastern Diffuse Region}
The column density of the large grains can be derived from the intensities at two bands longer than or equal to 100 $\mu$m under the assumption of a single-temperature model (e.g., \cite{Shibai1999}).
Here, the optical depth at 100 $\mu$m ($\tau_{100}$) is derived from the IRAS 100 $\mu$m and the FIRBE 155 $\mu$m intensities by using the method described in subsection 4.2.
The resultant optical depth, $\tau_{100}$, ranges from 0.15 to 0.0015 in the region above the 2.6 $\sigma$ intensity level in the FIRBE 155 $\mu$m map (figure 6a).
On the contrary, the dust optical depth derived from the IRAS 60 and 100 $\mu$m bands in the same manner ($\tau_{\mathrm{IRAS}}$) is 5 -- 300 times smaller than the optical depth, $\tau_{100}$, mentioned above.
This discrepancy can be attributed to that $\tau_{\mathrm{IRAS}}$ does not directly represent the mass of the large grains because the IRAS 60 $\mu$m intensity is not dominated by the large grains, but by the very small grains. 
Therefore, the column density, $\tau_{\mathrm{IRAS}}$, and the dust mass derived from the IRAS 60 and 100 $\mu$m intensities would be 5 -- 300 times less than the actual values for the large grains.

Next, we derive the gas column density, and compare it with that obtained by CO observations.
Assuming the gas-to-dust ratio ($R$) is 100, the gas column density ($N_\mathrm{H}$) is obtained by the following equation:
\begin{equation}
N_{\mathrm{H}} = \frac{R\ \tau_{100}}{m_{\mathrm{H}}\ \kappa_{100}}, 
\end{equation}
where $\kappa_{100}$ is the mass-absorption coefficient at 100 $\mu$mm. 
We use 29.3 cm$^{2}$ g$^{-1}$ for it according to the semi-empirical value by \citet{Hildebrand1983}. 
This value is also consistent with the value of 30.0 cm$^{2}$ g$^{-1}$ derived from the result of the laboratory measurement by \citet{Agladze1996}.
The thus-derived gas column density map is shown in figure 6a.
For a comparison, figure 6b shows a ${}^{13}$CO (J = 1--0) total integrated intensity map with $\timeform{3'}$ spatial resolution \citep{Nagahama1998}, where the velocity range is from 0 to 15 K km s$^{-1}$.
The integral-shaped filament (ISF), found by a molecular gas observation, is clearly seen in figure 6a: the high column density ridge extends from north through Orion KL to south except for the central rectangular blank region where the IRAS 100 $\mu$m is saturated. 
Actually, the SCUBA observation at 450 and 850 $\mu$m by \citet{Johnstone1999} reveals that this ISF ridge is common for the CO molecular gas and the dust emission.

In figure 6a, one can see the dense dust distribution surrounding the small H\emissiontype{II} regions, L 1641-N and NGC 1999 at the southern edge of ISF (declination is from $\timeform{-6.2D}$ to $\timeform{-7.0D}$), similar to the CO gas distribution.
In the region where R.A. = $\timeform{5h36.2m} \sim \timeform{5h37.0m}$ and DEC. = $\timeform{-6.2D} \sim \timeform{-6.5D}$, the column density derived from the dust emission (figure 6a) is significantly small due to the spuriously enhanced IRAS 100 $\mu$m emission.
Except for this region, the gas column density map derived from the dust emission (figure 6a) represents the actual distribution of the interstellar matter in each line of sight more precisely than that derived from the CO molecular lines because the ambiguity in the gas-to-dust ratio is smaller than that in the total gas to the CO gas ratio.

\subsection{The Gas-to-Dust Ratio}
Here, we derive the gas-to-dust mass ratio (exactly the hydrogen gas mass estimated from CO observation to the large grain mass ratio) along the ISF according to the above discussion.
The gas mass is calculated by using an equation in \citet{Nagahama1998}.
The dust mass $(M_{\mathrm{dust}})$ is calculated from the equation, $M_{\mathrm{dust}} = (\tau_{100}/\kappa_{100}) D^2 \Omega$, where $\Omega$ is the solid angle of the beam and D is the distance to Orion KL (470 pc).
The compared area is the region along the ISF and the south-southeastern diffuse region that have the 155 $\mu$m intensity above the lowest contour in figure 6a.
The optical depth at 100 um, $\tau_{100}$, ranges from 0.015 to 0.15 in this compared area.
Figure 7a shows a gray map of the dust superposed on the ${}^{13}$CO integrated intensity map.
The other panels in figure 7 also show the distributions of the dust temperature, dust mass, the gas mass, and the gas-to-dust ratio.
It is seen that the gas-to-dust ratio gradually decreases around the northern edge of ISF. 
The possible reason for this is that the UV photons originated by the NGC 1977 ionized region penetrate and dissociate the CO molecular gas.
In the region between $\timeform{+20'}$ and $\timeform{-20'}$ of the declination offset, the gas-to-dust ratios are in the range from 100 to 200.
This value is consistent with the mean value of 160 $\pm$ 60 for the galactic plane obtained from COBE observation \citep{Sodroski1994}.
On the other hand, the gas-to-dust ratio ranges from 30 to 100 for the southern area, less than $\timeform{-30'}$ of the declination offset.
The dust mass distribution shows the trend of gradually increasing toward the southern direction and the mass concentration around $\sim \timeform{40'}$.
The averaged dust temperature for this region shows a low value around 16 K, which is similar to the mean dust temperature in the outer galactic plane \citep{Shibai1999, Sodroski1994}.
The lower gas-to-dust ratio and the lower dust temperature suggest the possibility that the CO molecular gas may be depleted in the icy grain mantles \citep{Thi2001}.
\citet{Thi2001} estimated the total gas masses by observing $H_2$ pure-rotational line of the disks around young stars. 
The ratios of the gas masses derived from CO observation to dust masses show factors of 10 -- 200 lower than the standard value of 100, because CO molecular gas is photodissociated and frozen onto grains in the cold dense part of disks.
On the contrary, the depletion of gas-to-dust ratio in cold dust region may be explained as follows.
One problem with a single temperature model as compared to a radiation transfer model is: in single temperature model, one basically gets the temperature of the dust of peak emission. 
Even if cold dust is present, its presence is not easily seen in the presence of warm dust, unless one has long wavelength observation. 
On the other hand, if warm dust is absent, cold dust is more easily inferred, even with two band data. Since a lower temperature gives a larger dust mass, one tends to underestimate the gas-to-dust ratio, and vice versa
Part of the variation seen in the gas-to-dust ratio may be due to this effect.
In addition, this lower temperature indicates that there is no active star-forming region in the southern edge of ISF and the south-southeastern diffuse region.
There are only smaller star-forming sources of L 1641-N and NGC 1999 and no bright IRAS sources.

\section{Conclusion}

This paper presents the result of a wide-area mapping of the 155 $\mu$m continuum emission in the Orion molecular cloud complex.
The Japanese balloon-borne telescope (FIRBE) surveyed a sky area over 50 deg$^2$ and detected far-infrared emission over the region of 1.5 deg$^2$ around the KL nebula with $\timeform{3'}$ resolution.
The peak emission at Orion KL has a flux value of 42.5 $\pm$ 4.8 kJy in the $\phi \ \timeform{3'}$ beam.
The far-infrared distribution of intense region is similar that of the IRAS 100 $\mu$m image.
The dust temperature and optical depth of the thermal-equilibrium large grain are estimated by applying a single-temperature dust model modified with a $\lambda^{-2}$ emissivity law to the IRAS 100 and FIRBE 155 $\mu$m intensities.
They range from 15 to 30 K and from 0.15 to 0.0015, respectively.
Due to the significant contribution of the statistically heated very small grain in the IRAS 60 $\mu$m intensity, the dust temperature T(60/100) derived from the IRAS 60 and IRAS 100 $\mu$m intensities is 5 -- 15 K higher than the thermal-equilibrium dust temperature.
As a result, the dust optical depth at 100 $\mu$m, derived from T(60/100) has a 5 -- 300 times smaller value.
Toward the H\emissiontype{II} region, M 42, the IRAS 60 $\mu$m and also IRAS 25 $\mu$m intensities are relatively stronger than those outside of the H\emissiontype{II} region.
This tendency in the H\emissiontype{II} region may be attributed to the result of the statistically heating of very small grains by double photon incidence \citep{Okumura1999}.
The region where the dust temperature is greater than 20 K around M 42 has an extent similar to that of the optical region.
This fact suggests that the far-infrared emission around M 42 is mainly provided by the Trapezium Cluster.
The optical-depth distribution shows a filamentary dust ridge that has a $\timeform{1.5D}$ extent in the north -- south direction.
The shape of this dense dust ridge resembles the Integral-Shaped Filament (ISF) molecular gas distribution.
The gas-to-dust ratio derived from the CO molecular gas distribution along the ISF shows 100 -- 200 around the M 42 ionized region on the ISF. However, the gas-to-dust ratio decreases in other regions of the ISF, which may be interpreted as being an effect of CO depletion due to the photodissociation and/or freezing on dust grains.

\bigskip
We would like to acknowledge Dr. T. Nagahama and Dr. Y. Fukui for allowing us to use the ${}^{13}$CO data and Dr. X. Dupac for providing us with the PRONAOUS data.
We also express our special thanks to the technical staff of Tata Institute of Fundamental Research and National Balloon Facility of India.
This work was supported by the Grant-in Aid of the Japan Society for the Promotion of Science (10147102) and by the Research Fellowship of Japan Society for the Promotion of Science.


\clearpage
\begin{figure}
\begin{center}
\FigureFile(80mm,80mm){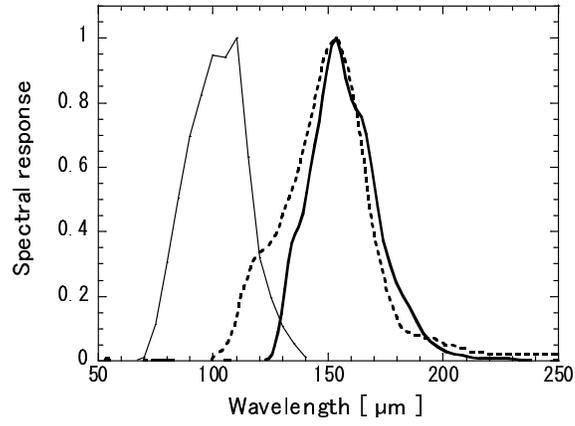}
\end{center}
\caption{
The thick curve shows the relative systematic spectral response of the FIRBE 155 $\mu$m band. The thin-curve and the broken-curve are the spectral responses of the IRAS 100 $\mu$m band and the COBE/DIRBE 140 $\mu$m band, respectively (IRAS and COBE/DIRBE Explanatory Supplement). All curves are normalized at their peaks.
}
\end{figure}

\clearpage
\begin{figure}
\begin{center}
\FigureFile(80mm,80mm){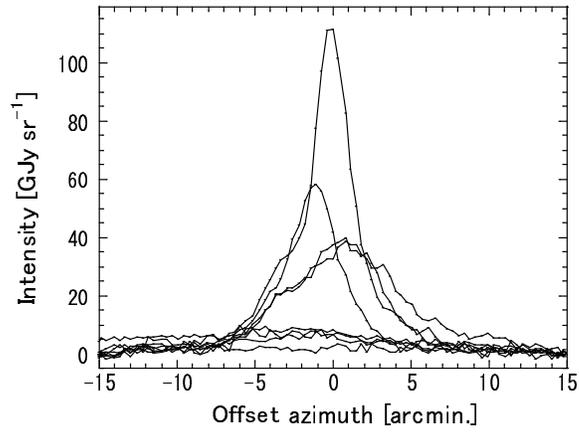}
\end{center}
\caption{
Example of raw detector signals of eight pixels in one column of the 4 by 8 pixel array at crossing Orion KL. The horizontal axis shows the azimuth offset relative to the peak position that corresponds to Orion KL.
}
\end{figure}

\clearpage
\begin{figure}
\begin{center}
\FigureFile(160mm,160mm){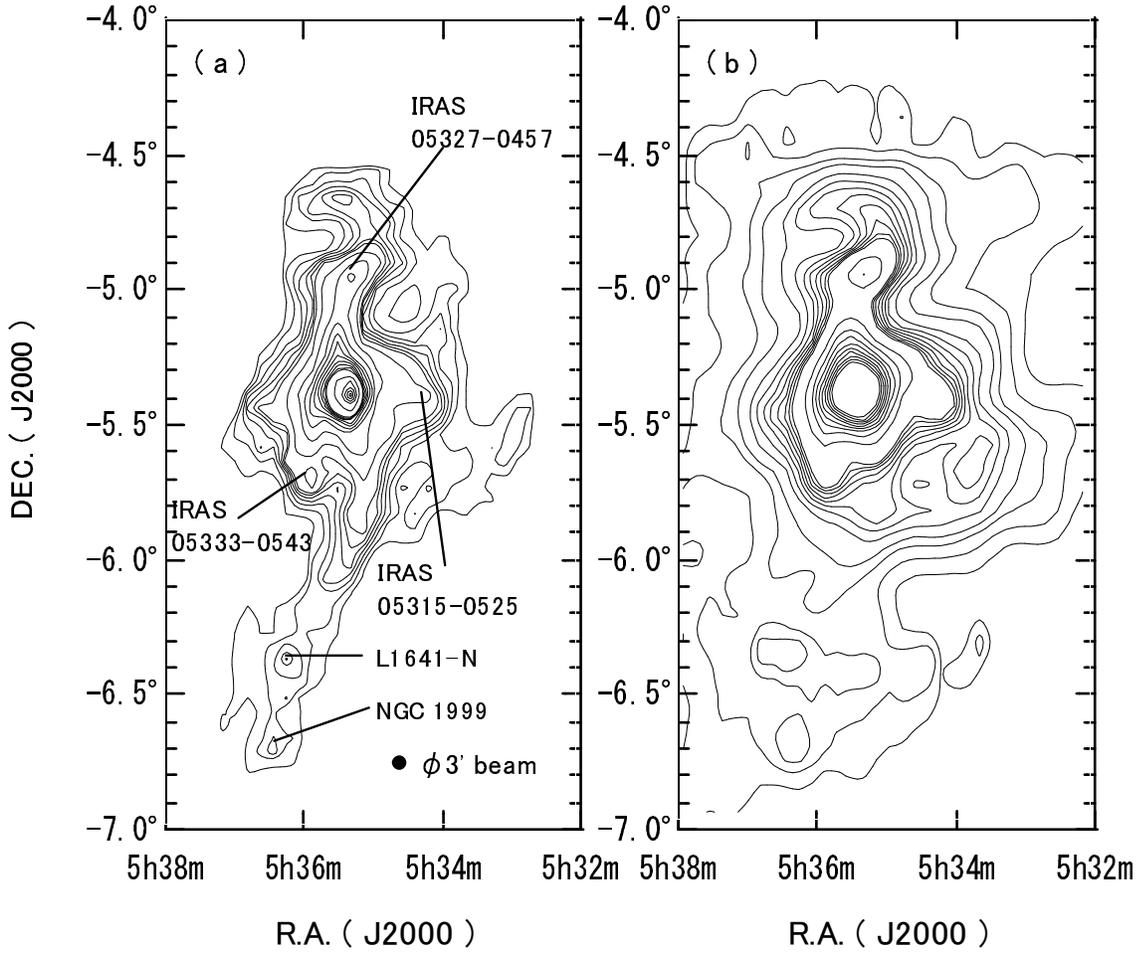}
\end{center}
\caption{
(a) 155 $\mu$m continuum intensity map observed by the FIRBE telescope. The spatial resolution is $\timeform{3'}$ and the intensity unit is MJy sr$^{-1}$. The lowest contour levels are 352 (=2.6 $\sigma$), 488 (= 3.6$\sigma$), 600, 700, 800, 900, 1000, 1300, 1000 interval up to 10000, 10000 interval up to peak intensity of 71080 MJy sr$^{-1}$ corresponding to the flux of 42.5 kJy in $\phi \ \timeform{3'}$ beam.
(b) The IRAS 100 $\mu$m intensity map of the same area after correction for the DIRBE -- IRAS calibration scale difference (multiplied by 0.72). The contour levels are 70, 100, 150, 100 interval from 200 up to 1000, and 1000 interval up to 10000. The IRAS 100 $\mu$m intensity around Orion KL is saturated and the sprious patterns appear at an angular distance around $\timeform{1D}$ away from Orion KL.
}
\end{figure}

\clearpage
\begin{figure}
\begin{center}
\FigureFile(160mm,160mm){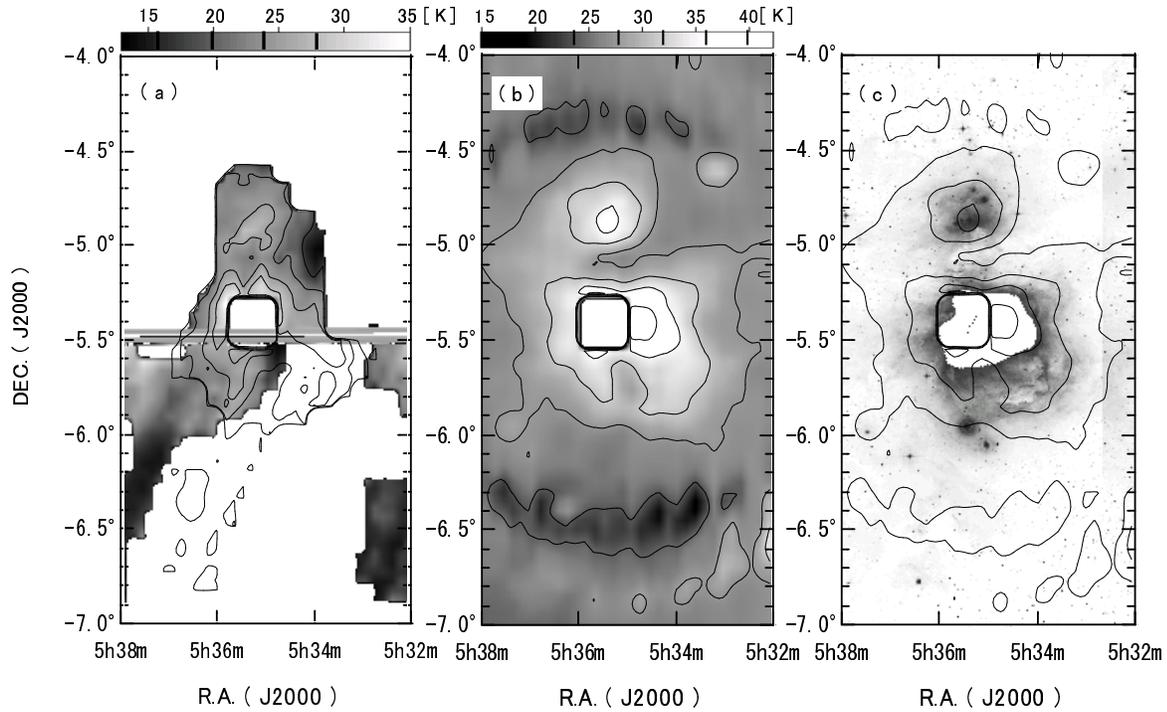}
\end{center}
\caption{
(a) Dust temperature ($T_{\mathrm{LG}}$) map derived from the 155 $\mu$m intensity of the present work and the IRAS 100 $\mu$m intensity. The contours indicate 16, 20, 24, and 28 K.
(b) Same as (a) from the IRAS 60 and 100 $\mu$m intensities.
The contours indicate at intervals of 4 K from 24 to 40 K.
(c) Red-band image of the Digital Sky Survey overlaid on the contour map of (b). The white-colored region represents the H\emissiontype{II} region determined from the 330 MHz radio continuum map by \citet{Subrahmanyan2001}.
}
\end{figure}

\clearpage
\begin{figure}
\begin{center}
\FigureFile(100mm,100mm){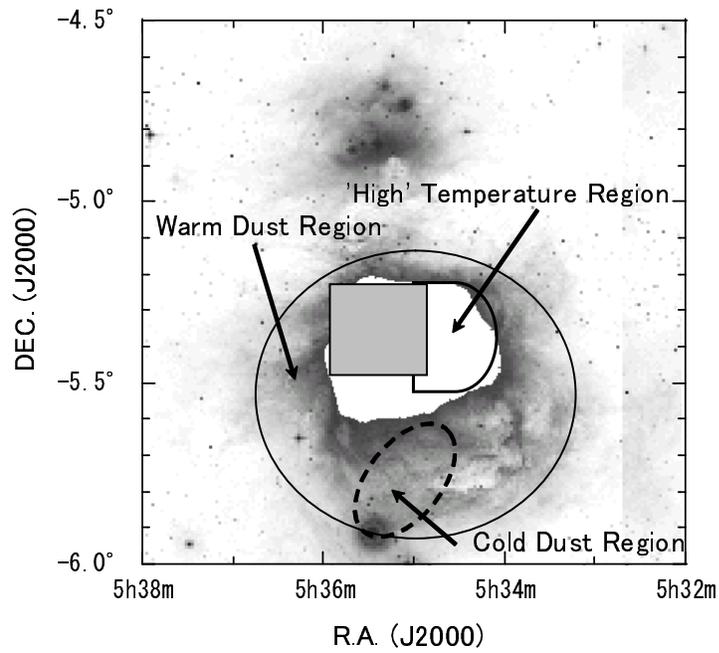}
\end{center}
\caption{
Schematic diagram for defining the three regions around M 42 superposed on the red-band image.
The gray central rectangular region indicates the region where the IRAS 100 $\mu$m band is saturated.
}
\end{figure}

\clearpage
\begin{figure}
\begin{center}
\FigureFile(160mm,160mm){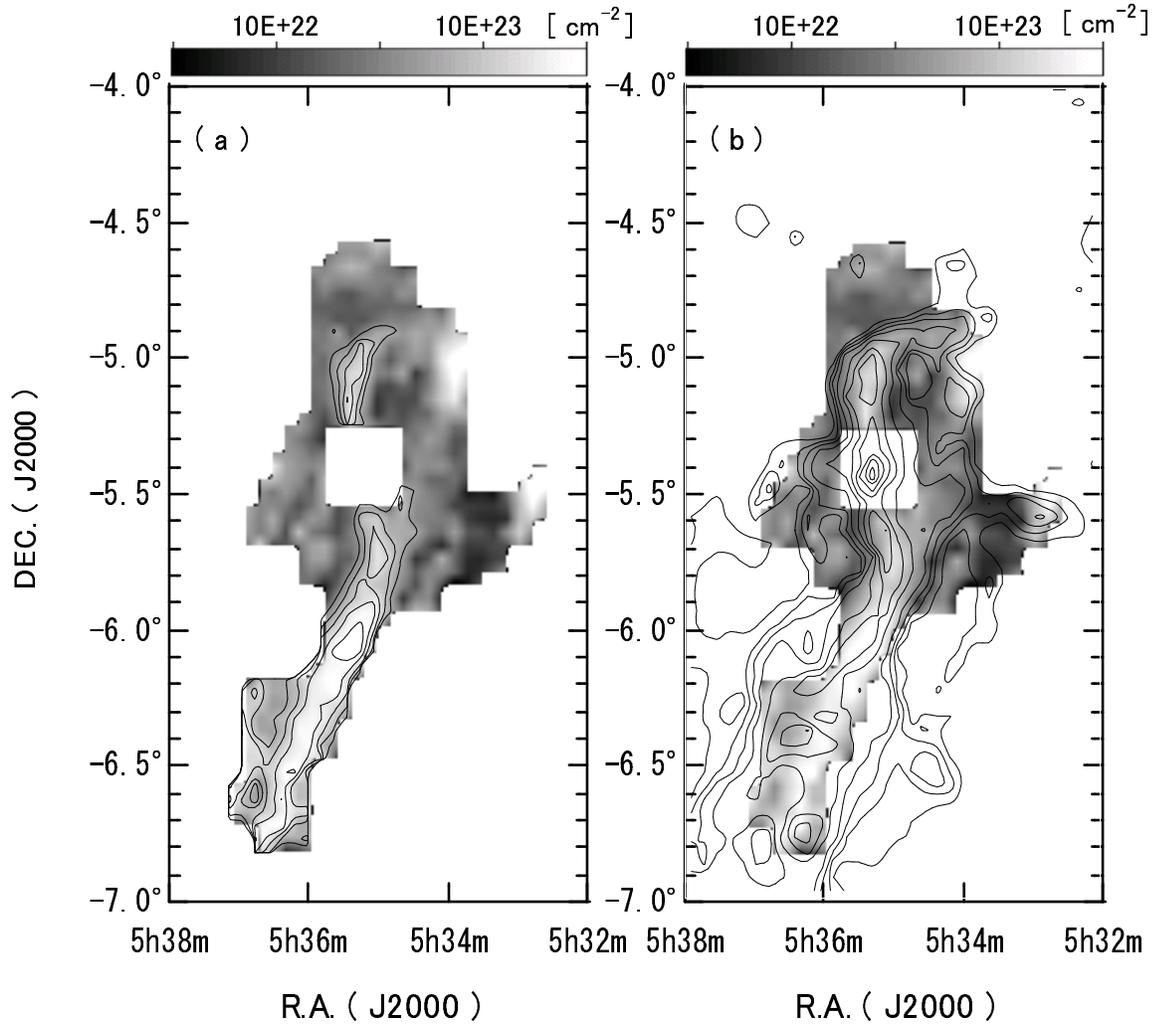}
\end{center}
\caption{
(a)Gas column density map derived from the dust optical depth and a gas-to-dust mass ratio of 100.
Logarithmic contours are shown along the ISF at the interval of 0.2 from 22.5 up to 23.5.
(b) ${}^{13}$CO (J = 1--0) total integrated intensity map with $\timeform{3'}$ spatial resolution \citep{Nagahama1998}, where the velocity range is from 0 to 15 K km s$^{-1}$, is overlaid on the color map described in figure 6a. 
Contour levels are at 3.0, 5.0, 7.0, and at the intervals of 5 K km s$^{-1}$ from 10.0 K km s$^{-1}$.
}
\end{figure}

\clearpage
\begin{figure}
\begin{center}
\FigureFile(160mm,160mm){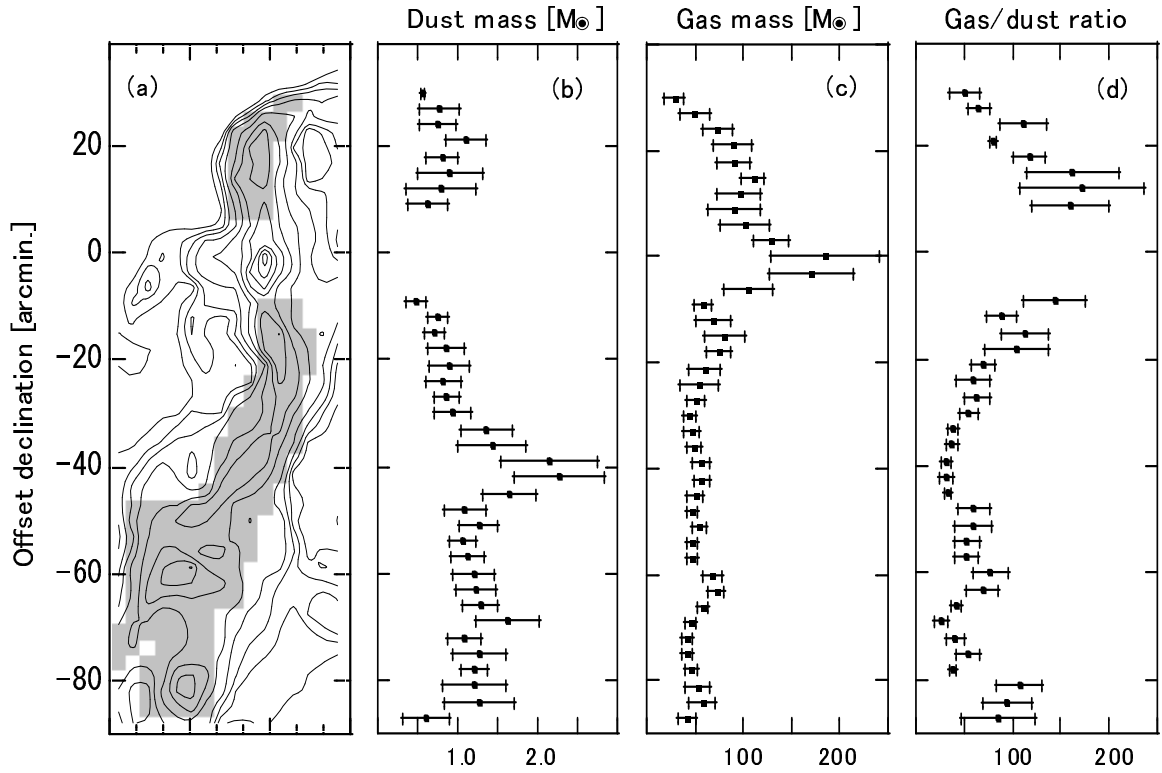}
\end{center}
\caption{
(a) The gray area indicates the area for which the dust mass was derived as in (b). The contour is the ${}^{13}$CO integrated intensity map. The vertical axis is the declination offset relative to the OMC-1 position. (b) Mean dust mass distribution along the ISF ridge, averaged in each $\timeform{3'}$ bin in declination. (c) Same as (b) for the gas mass. (d) Gas-to-dust distribution.
}
\end{figure}


\clearpage
\begin{table}
\caption{Intensity ratio of the `High' Temperature Region to the surrounding reference region.}
\begin{center}
\begin{tabular}{lccccc}
\hline
\hline
 & \multicolumn{4}{c}{\textit{IRAS}} & \textit{FIRBE} \\
\cline{2-6}
Band & 12 $\mu$m & 25 $\mu$m & 60 $\mu$m & 100 $\mu$m & 155 $\mu$m \\
\hline
Intensity ratio & 1.30 $\pm$ 0.20 & 1.72 $\pm$ 0.42 & 
1.71 $\pm$ 0.36 & 1.26 $\pm$ 0.21 & 1.25 $\pm$ 0.25 \\
\hline
\end{tabular}
\end{center}
\end{table}

\clearpage



\begin{thebibliography}{}

\bibitem[Agladze \etal (1996)]{Agladze1996} 
  Agladze, N., I., Sievers, A., J., Jones, S. A., Burlitch, J., M., \&
  Beckwith, S., V., W. 1996, \apj, 462, 1026
\bibitem[Arimura \etal (2000)]{Arimura2000}
   Arimura, S., \etal,2000, \procspie, 4014, 237
\bibitem[Bally \etal (1991)]{Bally1991} 
   Bally, J., Langer, W. D., \& Liu, W. 1991, \apj, 383, 645
\bibitem[Bally \etal (1987)]{Bally1987}   
   Bally, J. S., Antony A., Wilson, R. W., \& Langer, W. D. 1987, \apj, 312, 45
\bibitem[Becklin \etal (1976)]{BN}
  Becklin, E. E., Neugebauer, G., Beckwith, S., Gatley, I., Matthews, 
  K., Sarazin, C., \& Werner, M. W.
  1976, \apj, 207, 770
\bibitem[Boulanger \etal (1996)]{Boulanger1996}
  Boulanger, F., Abergel, A., Bernard, J. P., Burton, W. B., 
  Desert, F. X., Hartmann, D., Lagache, G., \& Puget, J. L.
  1996, \aap, 312, 256
\bibitem[Brown \etal (1994)]{Brown1994}
  Brown, A. G. A., de Geus, E. J., \& de Zeeuw, P. T.
  1994, \aap, 289, 101
\bibitem[Cesanori and Wilson (1994)]{Cesanori1994}   
  Cesanori, R., \& Wilson, T. L. 1994, \aap, 281, 209
\bibitem[Chini \etal (1997)]{Chini1997} 
  Chini, R., Reipurth, Bo, Thompson, W., Bally, J., Nyman, L.-A., Sievers, A. 
  \& Billawala, Y. 1997, \apj, 474, L135
\bibitem[Doi \etal (2000a)]{Doi2000-a} 
  Doi, Y., \etal 2000a, Advances in Space Research, 25, 11, 2285
\bibitem[Doi \etal (2000b)]{Doi2000-b}
  Doi, Y., \etal 2000b, Experimental Astronomy, 10, 393
\bibitem[Dupac \etal (2001)]{Dupac2001}
  Dupac, X., \etal 2001, \apj, 553, 604
\bibitem[Dutrey \etal (1991)]{Dutrey1991}
  Dutrey, A., Langer, W. D., Bally, J., Duvert, G.,
  Castets, A., \& Wilson, R. W. 1991, \aap, 247, L9
\bibitem[Genzel and Stutzki (1989)]{Genzel1989}  
  Genzel, R., \& Stutzki, J. 1989, \araa, 27, 41
\bibitem[Gordon (1988)]{Gordon1988} 
  Gordon, M. A. 1988, \apj, 331, 509
\bibitem[Goudis (1982)]{Goudis1982}  
   Goudis, C. 1982, British Astron. Assoc. Circ., 93, 45
\bibitem[COBE/DIRBE Explanatory Supplement]{DIRBE}
  Hauser, M. G., Kelsall, T., Leisawitz, D., \& Weiland, J (ed.) 1993, 
  COBE Diffuse Infrared Background Experiment (DIRBE) Explanatory Supplement 
\bibitem[Hildebrand (1983)]{Hildebrand1983}  
  Hildebrand, R. H. 1983, \qjras, 24, 267
\bibitem[Ikeda \etal (2002)]{Ikeda2002}
  Ikeda, M., Oka, T., Tatematsu, K., Sekimoto, Y., \& Yamamoto, S. 
  2002, \apjs, 139, 467
\bibitem[Johnstone and Bally (1999)]{Johnstone1999} 
  Johnstone, D. \& Bally, J. 1999, \apj, 510, L49
\bibitem[Kleinmann, Low (1967)]{KL}
  Kleinmann, D. E., \& Low, F. J. 1967, \apj, 149, L1
\bibitem[Lis \etal (1998)]{Lis1998}  
  Lis, D. C., Serabyn, E., Keene, J., Dowell, C. D., Benford, D. J., 
  Phillips, T. G., Hunter, T. R., \& Wang, N. 1998, \apj, 509, 299
\bibitem[Mookerjea \etal (2000)]{Mookerjea2000-a} 
  Mookerjea, B., Ghosh, S. K., Rengarajan, T. N., Tandon, S. N.,
  \& Verma, R. P. 2000, \aj, 120, 1954
\bibitem[Nagahama \etal (1998)]{Nagahama1998}  
  Nagahama, H., Mizuno, A., Ogawa, H., \& Fukui, Y. 
  1998, \aj, 116, 336
\bibitem[Nagata \etal (2002)]{Nagata2002}  
  Nagata, H., Shibai, H., Takeuchi, T., T., \& Onaka, T. 
  2002, \pasj, 54, 695
\bibitem[O'Dell (2001)]{O'Dell2001}
   O'Dell, C. R. 2001, \araa, 39, 99
\bibitem[Okumura \etal (1999)]{Okumura1999}
  Okumura, K., Hiromoto, N., Shibai, H., Onaka, T., Makiuti, S., 
  Matsuhara, H., Nakagawa, T., \& Okuda, H.
  1999, in Star Formation 1999, ed.\ Nakamoto, T. 
  (Nobeyama: Nobeyama Radio Observatory), 96
\bibitem[Ristorcelli \etal (1998)]{Ristorcelli1998}
  Ristorcelli, I., \etal 1998, \apj, 496, 267
\bibitem[Sakamoto \etal (1994)]{Sakamoto1994} 
  Sakamoto, S., Hayashi, M., Hasegawa, T., Handa, T., \& Oka, T. 
  1994, \apj, 425, 641
\bibitem[Shibai \etal (2002)]{Shibai2002}    
  Shibai, H., \etal 2002, Advances in Space Research, 30, 5, 1289
\bibitem[Shibai \etal (1999)]{Shibai1999}
  Shibai, H., Okumura, K., \& Onaka, T. 
  1999, in Star Formation 1999, ed. T. Nakamoto 
  (Nobeyama: Nobeyama Radio Observatory), 67
\bibitem[Sodroski \etal (1994)]{Sodroski1994}
  Sodroski, T., J., \etal 1994, \apj, 428, 638
\bibitem[Subrahmanyan \etal (2001)]{Subrahmanyan2001}  
  Subrahmanyan, R., Goss, W. M., \& Malin, D. F.
  2001, \aj, 121, 399
\bibitem[Tatematsu \etal (1993)]{Tatematsu1993}  
  Tatematsu, K., \etal 1993, \apj, 404, 643
\bibitem[Thi \etal (2001)]{Thi2001}
  Thi, W.-F., van Dishoeck, E. F., Blake, G. A., 
  van Zadelhoff, G.-J., \& Hogerheijde, M. R. 
  1999, \apj, 521, L63
\bibitem[Wall \etal (1996)]{Wall1996}
  Wall, W. F., \etal 1996, \apj, 456, 566

\end{thebibliography}
\end{document}